# Computational Study of Molecular Mechanisms of Caffeine Actions.

V. I. Poltev, E. Rodríguez, T. I. Grokhlina, A. V. Teplukhin, A. Deriabina, and E. González

*Abstract*— Caffeine (CAF) is one of the most widely and regularly consumed biologically active substances. We use computer simulation approach to the study of CAF activity by searching for its possible complexes with biopolymer fragments. The principal CAF target at physiologically important concentrations refers to adenosine receptors. It is a common opinion that CAF is a competitive antagonist of adenosine. At the first step to molecular level elucidation of CAF action, we have found a set of the minima of the interaction energy between CAF and the fragments of human A1 adenosine receptor. Molecular mechanics is the main method for the calculations of the energy of interactions between CAF and the biopolymer fragments. We use the Monte Carlo simulation to follow various mutual arrangements of CAF molecule near the receptor. It appears that the deepest energy minima refer to hydrogen-bond formation of CAF with amino acid residues involved in interactions with adenosine, its agonists and antagonists. The results suggest that the formation of such CAF-receptor complexes enforced by a close packing of CAF and the receptor fragments is the reason of CAF actions on nervous system. CAF can block the atomic groups of the adenosine repressors responsible for the interactions with adenosine, not necessary by the formation of H bonds with them, but simply hide these groups from the interactions with adenosine.

## I. INTRODUCTION

Caffeine (CAF) acts as a stimulant of the nervous system [1]. This is the reason for consumption by the most humans of various CAF-containing drugs and beverages. Several viewpoints exist on CAF consumption supported by experimental observations, starting from a declaration that "caffeine produces special beneficial effects for the mind, body, and spirit" [2], to "CAF can produce a clinical dependence syndrome and has a potential for abuse" [3]. Besides its principal and rapid actions on nervous system,

Manuscript received March 30, 2010. This work was supported in part by the Autonomous University of Puebla under project 29/EXC/08I.

V. I. Poltev is with the Department of Physics and Mathematics, Autonomous University of Puebla, Puebla 72570, Mexico (phone: 52-222-229-5500; fax:52-222-229-5636; e-mail:poltev@fcfm.buap.mx); and with the Institute of Theoretical and Experimental Biophysics, Pushchino, Moscow Region 142290, Russia.

E. Rodríguez, A. Deriabina, and E. González are with the Department of Physics and Mathematics, Autonomous University of Puebla, Puebla 72570, Mexico (e-mail: gonzalez@fcfm.buap.mx).

T. I. Grokhlina, and A. V. Teplukhin are with the Institute of Mathematical Problems of Biology, Pushchino, Moscow Region 142290, Russia (e-mail: tepl@impb.psn.ru).

CAF has numerous effects on various other biological processes, including DNA functioning and enzyme activity.

The beginning of our computational studies of the molecular mechanisms of CAF actions refers to the consideration of its possible interactions with DNA fragments in relation to "lateral" action on genetic processes (e.g. [4], [5] and references therein). Computer simulation enables us to explain CAF influence on DNA repair and DNA-drug interactions. Recently [6] we started our preliminary attempts to understand the principal action of this important compound with rather simple molecular structure (see Fig. 1, center).

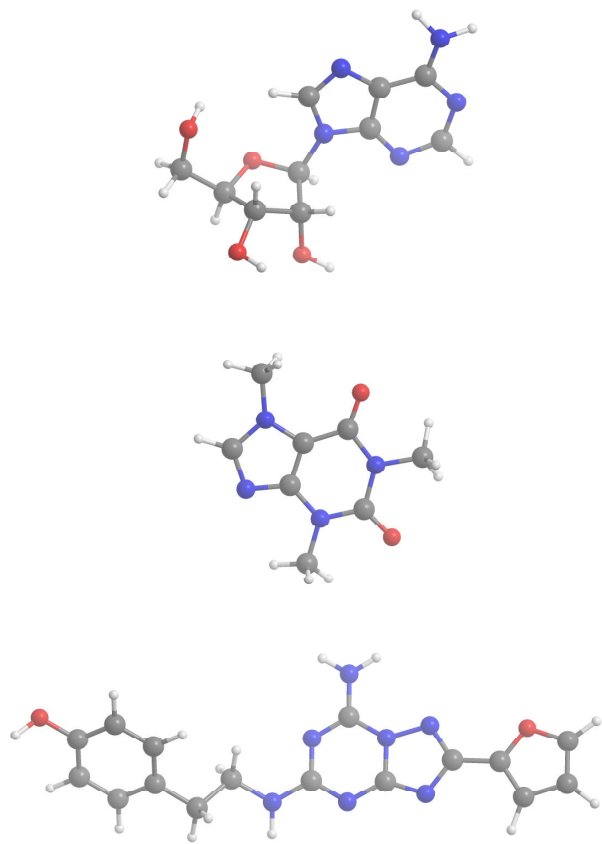

Fig. 1. The molecular structure of adenosine (top), caffeine (center), and ZM241385 antagonist (bottom).

The main CAF target at physiologically important concentrations refers to membrane proteins, namely to A1 and A2a adenosine receptors [7]–[9]. Adenosine itself (Fig. 1, top) is an important biomolecule, and it is a part of many other biomolecules, including nucleic acids and coenzymes. Extracellular adenosine plays an essential role in physiology and initiates its effects through the activation of adenosine receptors. It is a common opinion that CAF is a competitive antagonist of adenosine.

In view of the important role of adenosine receptors, a series of adenosine agonists and antagonist have been synthesized and studied (see e.g. [10], [11]). All of them have a more complex molecular structure than CAF and adenosine itself, and more possibilities for complex formations with the receptors. The molecular structure of one of adenosine antagonists, ZM241385, is displayed in Fig. 1, bottom. The only quantitative experimental data on adenosine receptor three-dimensional structure refer to ZM241385-A2a receptor complex [12]. It is a challenge to understand "Why such rather small molecule as CAF with restricted set of centers of sufficiently strong interactions can compete with larger molecules capable to form more H bonds, and hence potentially more stable complexes?" We started a search for an answer to this question from the computational study of possible complexes of CAF with short fragments of A1 receptor [6]. Here we consider CAF complexes with a longer receptor fragment, and we come to some general conclusions on CAF actions.

## II. MOLECULAR STRUCTURE OF CAFFEINE AND ITS POSSIBLE INTERACTIONS WITH BIOMOLECULES.

The structure of the CAF molecule is relatively simple (see Fig. 1, center). CAF is a purine derivative, 1,3,7-trimethylxantine (1,3,7-trimethyl-2,6-dioxopurine). It contains three hydrophilic centers, namely, hydrogen-bond acceptor atoms O2, O6, and N9, and three methyl groups, and it has no proton-donor groups. All the atoms except hydrogens of the methyl groups pertain to the same plane. A combination of hydrophilic and hydrophobic atomic groups enables CAF to be soluble in both polar (water) and nonpolar solvents. These properties of CAF allow its passage through all biological membranes. The absence of proton-donor groups results in a reduction of number of possible stable configurations of CAF complexes with the adenosine receptors as well as with the most of biomolecules. Such configurations can be formed via hydrogen bonds of CAF proton acceptors with NH groups of peptide chain and with proton-donor groups of some amino-acid residues (Ser, Thr, Asn, Gln, Tyr, His, and Trp). Another possibility of CAF interactions with proteins refers to stacking interactions with aromatic amino acid residues. In this paper we consider possible interactions of CAF with a fragment of adenosine A1 receptor containing 9 amino acid residues.

Both adenosine and its agonists and antagonists have more centers for interactions with proteins, e. g., adenosine has 7 proton acceptor groups and 6 hydrogens capable to participate in H bonds (see Fig. 1). The ZM241385 antagonist has 5 proton acceptor groups and 4 hydrogens capable to form H bonds. Besides, both Adenosine and its agonists and antagonists are flexible molecules; the rotations about the single bonds are possible. This flexibility enables more close contacts with the receptor fragments, while the only flexibility of CAF originates from rotations of methyl hydrogens around N-C bonds.

## III. ADENOSINE RECEPTORS AND THEIR POSSIBLE INTERACTIONS WITH CAFFEINE.

Four adenosine receptors have been characterized from several mammalian species including human beings: A1, A2a, A2b, A3 (see e.g. [7], [13]). All these receptors are proteins containing more than 300 amino acid residues. They are membrane proteins, and there is only one X-ray work [12] on three-dimensional structure of the A2a receptor in complex with its antagonist ZM241385 (the molecular structure of the antagonist is displayed in Fig. 1, bottom). The molecular models of the receptors have been constructed theoretically, but it is difficult to approve the models via direct experimental data.

The receptors include extracellular, intracellular, and transmembrane domains. Extracellular domain contains a chain of several amino end residues and three loops; intracellular domain consists of three loops and a chain of carbonyl end amino acid residues. Seven transmembrane domains contain about 20 amino-acid residues each, arranged in α-helix conformation. The major part of amino-acid residues in these domains refers to hydrophobic ones while some hydrophilic residues of the third and seventh domains are of primary importance for interactions with adenosine, its agonists, and antagonists [11].

In this paper we present the results of the consideration of CAF interactions with the fragment of the seventh transmembrane domain of A1 adenosine repressor. The fragment, Ala-Ile-Phe-Leu-Thr-His-Gly-Asn-Ser, contains four residues with proton-donor groups (Thr, His, Asn, Ser) and two aromatic (Phe, His) residues. We selected this fragment as its primary structure enables us to suggest the formation of complexes with CAF via both H bonding and stacking interactions.

## IV. COMPUTATIONAL METHODS.

The system considered here contains CAF molecule and the receptor fragment selected in the previous section. The molecular mechanics method was used for computation and search of the energy minima of the systems. Two computational programs were employed: version 9 of AMBER program package [14] and our own program specially designed for computations of complexes of simple aromatic ligands with protein fragments. We use this program earlier for consideration of possible CAF complexes with small peptide fragments including short fragments of A1 adenosine receptor [6]. The program suggests the construction of protein fragment using atomic building blocks with fixed bond lengths and valence angles. The methyl groups were placed at the amino and carboxyl ends of the fragment to avoid a formation of additional charged or proton donor groups. The quantitative results of possible complex geometries presented below were finally obtained with our own program, while the AMBER program was used to obtain preliminary data and to demonstrate the independence of the main conclusions from the details of the molecular mechanics method. The comparison of the complex structures obtained via two methods demonstrates nearly the same geometry. Some H-bond lengths in the complexes obtained via the AMBER program were a bit shorter than those obtained via our program.

The molecular mechanics method suggests an additivity of several energy terms, including those describing changes of bond lengths, of valence and torsion angles, as well as pairwise interactions between all atoms which are not bonded between themselves or with the other common atom. Our program calculates terms related to changes of torsion angles and to atom–atom interactions. Each atom–atom term consists of a Coulomb term and a Lennard–Jones 6-12 term, commonly used in molecular mechanics calculations (1). For description of interactions of hydrogen atoms capable of forming hydrogen bonds with potential proton acceptors, the 6-12 term is substituted by a 10-12 term (2).

$$E_{ij}=ke_ie_j/r_{ij} - A_{ij}/r_{ij}^{6} + B_{ij}/r_{ij}^{12} \quad (1)$$
$$E_{ij}=ke_ie_j/r_{ij} - A_{ij}^{(10)}/r_{ij}^{10} + B_{ij}^{(10)}/r_{ij}^{12} \quad (2)$$

In these equations $k$ is the numerical constant, $r_{ij}$ is the distance between atoms $i$ and $j$, $e_i$ and $e_j$ are effective charges on atoms $i$ and $j$. The coefficients $A_{ij}$, $B_{ij}$, $A_{ij}^{(10)}$, and $B_{ij}^{(10)}$ are adjustable parameters, whose numerical values are the same as in our recent papers (e.g. [4] - [6]). The energy is calculated and minimized as a function of variables, describing mutual position of CAF and peptide, rotations of CAF methyl groups around the bonds connecting them to the purine ring, and rotations around single bonds in amino acid residues. The parameters of torsion terms of the Kollman's force field [15] were employed without any changes. Standard gradient minimization techniques are used throughout the work.

Minimizations have been started from various mutual molecule positions, corresponding to both suggested complex configurations (as described in the previous section) and selected instantaneous configurations of Monte Carlo sampling (see below). Penalty functions were used in some cases to maintain H-bond geometry at preliminary steps of the minimization. The torsion angles of the peptide backbone were fixed in α-helix configuration. The calculations with the AMBER program were performed using force field of the paper [15].

Monte Carlo simulation enables one to evaluate mean values of energy and structural characteristics of a system under consideration. Here we use the Monte Carlo simulation to follow various mutual arrangements of CAF molecule near the receptor. A statistically significant sampling of configurations was obtained by Metropolis et al. algorithm [16]. CAF molecule was displaced by a random vector and all the variable torsion angles of CAF and amino-acid residues were randomly changed. Each new configuration with the energy E1 is accepted if the difference of energies ΔE = E1-E0 is negative or, when ΔE>0, if exp(-ΔE/RT)≥s [where s is a random number distributed in the interval (0,1)]. Otherwise, the former configuration is taken once again. The maximum shift and rotation values are chosen in such a way that no more than 50% of new configurations were rejected. A sequence of such multiple trials forms a Markov chain, the probability of appearance of a configuration is proportional to the Boltzmann factor exp(-ΔE/RT). Each of the configurations accepted by Metropolis procedure corresponds to an instantaneous structure of the system. Starting from one of the configurations with two H-bonds and using cluster approximation we obtained a series of possible mutual CAF-receptor fragment arrangements at ambient temperature (300K). The use of Monte Carlo sampling provides us with the most populated by CAF regions of the fragment surface and displacements of CAF molecule along the peptide.

## V. THE MINIMA OF INTERACTION ENERGY BETWEEN CAF AND THE RECEPTOR FRAGMENT

The search of interaction energy minima of the system containing CAF molecule and the fragment of seventh transmembrane domain of A1 receptor reveals a set of local energy minima. Here we present the results for three of them as examples. The minimum energy configurations are characterized by close, but not shortened, contacts between CAF methyl groups and hydrophobic groups of the fragment. Thus, CAF molecule protects part of the receptor atomic groups from the interactions with other biomolecules.

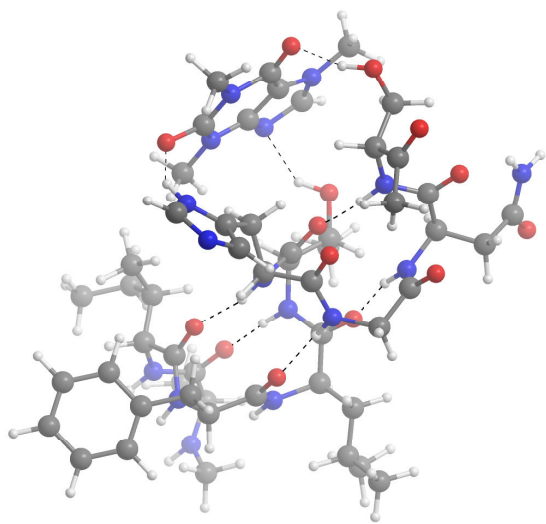

Fig. 2. The minimum energy configuration of the complex of CAF molecule (located in the top of the figure) with the fragment of the receptor. Atoms and bonds closer to observer are richer in color. Three H bonds of CAF with amino acid residues, as well as H bonds of peptide backbone of the α helix are shown as dashed lines.

The first minimum (Fig. 2) corresponds to the formation of three H bonds between CAF acceptors and amino-acid residues of the (Ala-Ile-Phe-Leu-Thr-His-Gly-Asn-Ser) fragment of the seventh domain of the A1 receptor. Three of four H-bond donors of the fragment participate in H bonding with CAF, namely N9(CAF)…H-O(Thr), O2(CAF)…H-N(His), and O6(CAF)…H-O(Ser). The second minimum (Fig. 3) refers to the formation of two H bonds between the CAF molecule and the receptor fragment, namely O2(CAF)…H-N(His), and O6(CAF)…H-O(Ser). This minimum is one of a few of minima with two CAF-fragment H bonds.

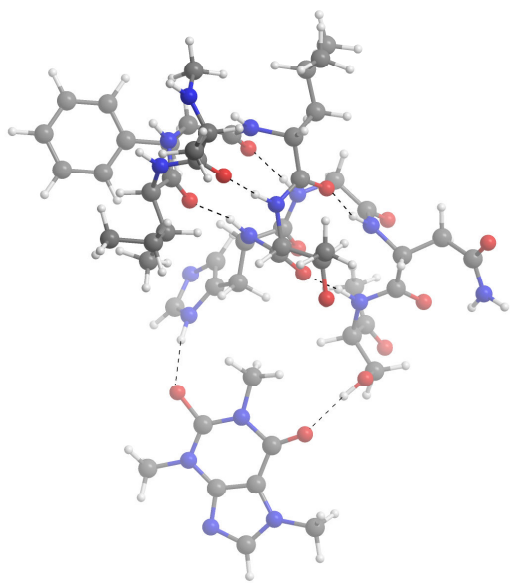

Fig. 3. The minimum energy configuration of the complex of CAF molecule (located in the bottom of the figure) with the receptor fragment. Two H bonds can be clearly seen. Designations are the same as in Fig. 2.

The third minimum considered (Fig. 4) corresponds to the single H-bond formation of CAF with the fragment and to the stacking arrangement of CAF and the His residue. The energy values and H-bond distances of the three minima are listed in Table 1. Two energy characteristics presented refer to the total energy change resulted from complex formation, ΔE, and the energy of interaction, $E_{int}$, between the CAF molecule and the receptor fragment in the configuration obtained.

TABLE I
ENERGY VALUES AND H BOND LENGTHS FOR THE THREE MINIMA

| Minimum No | 1 (Fig. 2) | 2 (Fig. 3) | 3 (Fig. 4) |
|---|---|---|---|
| ΔE (kcal/mol) | -20.49 | -18.55 | -21.59 |
| $E_{int}$ (kcal/mol) | -27.36 | -16.44 | -17.05 |
| 1st H bond | O6…H-O(Ser) | O6…H-O(Ser) | O2…H-O(Ser) |
| O…O; O…H (Å) | 2.77; 1.85 | 2.81; 1.85 | 2.78; 1.84 |
| 2d H bond | O2… H-N (His) | O2… H-N (His) | |
| O…N; O…H (Å) | 2.93; 1.93 | 2.88; 1.96 | |
| 3d H bond | N9… H-O (Thr) | | |
| N…O; N…H (Å) | 2.74; 1.94 | | |

The interaction energy can be greater or smaller than the total energy change. The first case refers to strong interaction between the CAF and the fragment (the first minimum with three H bonds). The formation of three H bonds results in less favorable residue configuration as compared to other minima with weaker CAF-fragment interactions. It is interesting, that the minimum with the single H bond has more negative both total and interaction energies as compared to the minimum with two H bonds.

VI. MONTE CARLO "TRACING" OF CAF MOVEMENTS.

The number of local minima of the energy CAF-receptor fragment system is enormously great; the configurations obtained via energy minimizations can approximately correspond to only minor part of CAF positions in live systems. To follow possible pathways of CAF movement and to search for starting configurations for the minimizations we performed Monte Carlo-Metropolis sampling for the system as described in Section IV.

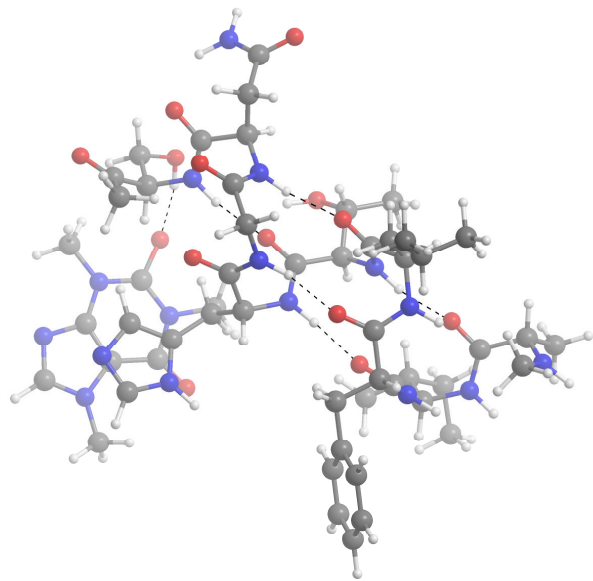

Fig. 4. The minimum energy configuration of the complex of CAF molecule (located in the left part of the figure) with the receptor fragment. The H bond of the CAF O2 atom with the Ser residue and the CAF-His stacking can be seen. Designations are the same as in Fig. 2.

Starting from the configuration corresponding to Minimum 2 (Fig. 3) we constructed the Markov chain of $16 \cdot 10^6$ steps. Each step includes a test change of all the variables of the system. The instantaneous configurations corresponding to every N-th step (N=n $\cdot 10^5$, where n is integer running from 1 to 160) were selected and analyzed. The possibilities of H-bond formations and stacking arrangements for all the configurations selected were checked. The variety of characteristics of the selected configurations was displayed at one-dimensional plot. The part of this plot corresponding to the first $55 \cdot 10^5$ steps (n varies from 0 till 55) is presented in the Fig. 5. This part corresponds to CAF movement near His residue of the fragment. The majority of the selected configurations of this part of the Markov chain correspond to H bonding of His N-H group to one of the three H-bond acceptors of the CAF molecule. The second H bond with Ser or Thr residues as well as stacking arrangement of CAF and Phe ring can form in some configurations. Only few configurations of this part and of the whole plot correspond to neither H bonds nor stacking arrangements.

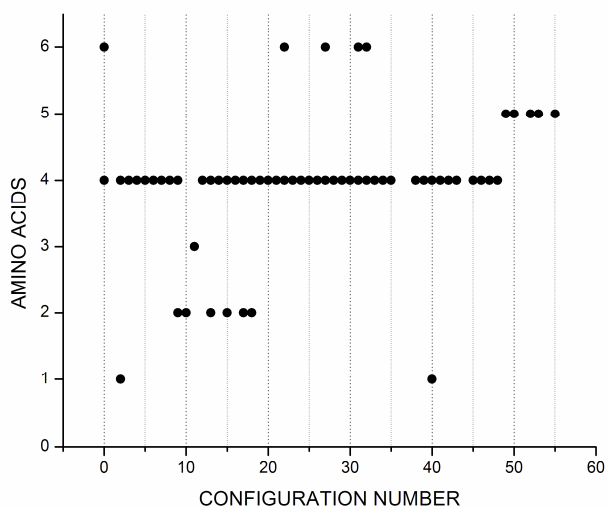

Fig. 5 The CAF positions at the receptor fragment in the selected configurations of Monte Carlo-Metropolis sampling. The dots correspond to H bonding or stacking arrangements of CAF and amino-acid residues; 1, Phe; 2, Thr; 3, His stacking; 4, His H bonding; 5, Asn; 6, Ser.

Other parts of the plot correspond to H bonding of CAF molecule to amino-acid residues or N-H groups of peptide backbone. The configurations with two H bonds, with single H bond, and with stacking arrangements occur in other parts as well.

## VII. CONCLUSIONS

Molecular mechanics calculations demonstrate an existence of energy minima for the system containing CAF and the fragment of the seventh transmembrane domain of the A1 adenosine receptor. The results suggest that the formation of CAF-receptor H-bonded complexes enforced by a close packing of CAF and the receptor fragments is the reason of CAF actions on nervous system. CAF can block the atomic groups of the adenosine repressors responsible for the interactions with adenosine, not necessary by the formation of H bonds with them, but simply hide these groups from the interactions with adenosine. The computer simulations make us closer to an answer to the question: "Why such rather small molecule as CAF with restricted set of the centers of sufficiently strong interactions can compete with larger molecules capable to form potentially more stable complexes?"